\documentclass[runningheads]{svmult}

\usepackage{makeidx} 
\usepackage{graphicx}
\usepackage{subeqnar}
\usepackage{multicol}
\usepackage{physprbb}
\usepackage[dvips]{color}
\usepackage{graphics}
\usepackage{epsfig}
\makeindex

\begin{document}

\newcommand{\greeksym}[1]{{\usefont{U}{psy}{m}{n}#1}}
\newcommand{\umu}{\mbox{\greeksym{m}}}
\newcommand{\udelta}{\mbox{\greeksym{d}}}
\newcommand{\uDelta}{\mbox{\greeksym{D}}}
\newcommand{\uPi}{\mbox{\greeksym{P}}}
\def\E{{\sl Einstein}}
\def\R{{\sl ROSAT}}
\def\A{{\sl ASCA}}
\def\C{{\sl Chandra}}
\def\X{{\sl XMM-Newton}}
\def\B{{\sl Beppo-Sax}}
\def\V{{\sl VLA}}

\def\gtap{\mathrel{\hbox{\rlap{\lower.55ex \hbox {$\sim$}}
                   \kern-.3em \raise.4ex \hbox{$>$}}}}
\def\ltap{\mathrel{\hbox{\rlap{\lower.55ex \hbox {$\sim$}}
                   \kern-.3em \raise.4ex \hbox{$<$}}}}
\newcommand{\K}{\mbox{\thinspace K}}
\newcommand{\pcm}{\mbox{cm$^{-3}$}}
\newcommand{\pcmsq}{\mbox{cm$^{-2}$}}
\newcommand{\ergsec}{\mbox{ergs s$^{-1}$}}
\newcommand{\ergcms}{\mbox{ergs cm$^{-2}$ s$^{-1}$}}
\newcommand{\kmsec}{\mbox{km s$^{-1}$}}
\newcommand{\Lsun}{\mbox{\rm\thinspace L$_{\odot}$}}
\newcommand{\Msun}{\mbox{M$_\odot$}}
\newcommand{\ml}{\mbox{\Msun\ yr$^{-1}$}}
\newcommand{\Lx}{\mbox{$L_{\rm x}$}}
\newcommand{\Fx}{\mbox{$F_{\rm x}$}}
\newcommand{\fx}{\mbox{$f_{\rm x}$}}
\newcommand{\uJy}{\mbox{$\mu$Jy}}
\newcommand{\boldf}{\mbox{\boldmath$f$\unboldmath}}
\newcommand{\boldfeq}[1]{\mbox{\boldmath$f = #1$\unboldmath}}
\newcommand{\chandra}{{\it Chandra}}
\newcommand{\chandraz}{{\it Chandra\/}}
\newcommand{\etal}{et al.}
\newcommand{\vmek}{\textsc{vmekal}}
\newcommand{\xspec}{\textsc{xspec}}
\def\fd{\hbox{$.\!\!^{\rm d}$}}
\def\fh{\hbox{$.\!\!^{\rm h}$}}
\def\fm{\hbox{$.\!\!^{\rm m}$}}
\def\fs{\hbox{$.\!\!^{\rm s}$}}
\def\fdg{\hbox{$.\!\!^\circ$}}
\def\farcm{\hbox{$.\mkern-4mu^\prime$}}
\def\farcs{\hbox{$.\!\!^{\prime\prime}$}}

\title*{X-Ray Supernovae}
\toctitle{X-Ray Supernovae}
\titlerunning{X-Ray Supernovae}

\author{Stefan Immler\inst{1}
\and Walter H.G.\ Lewin\inst{2}}
\authorrunning{Immler \& Lewin}

\institute{Astronomy Department, University of Massachusetts, Amherst, MA 01003
\and Center for Space Research and Department of Physics, \\
Massachusetts Institute of Technology,
Cambridge, MA 02139-4307}

\maketitle

\begin{abstract}
We present a review of X-ray observations of supernovae (SNe).
By observing the ($\sim0.1$--$100$~keV) X-ray emission from young SNe, 
physical key parameters such as the circumstellar matter (CSM) density, 
mass-loss rate of the progenitor and temperature of the outgoing and 
reverse shock can be derived as a function of time.
Despite intensive search over the last
$\sim25$ years, only 15 SNe have been detected in X-rays.
We review the individual X-ray observations of these SNe and
discuss their implications as to our understanding of the
physical processes giving rise to the X-ray emission.
\end{abstract}

\section{Introduction}

To date, several thousand supernovae (SNe) have been discovered
in the optical, whereas only $\sim$30 SNe have been detected in the 
radio and 15 in the X-ray band ($\sim0.1$--$100$~keV).
Neutrinos have been recorded from one nearby supernova only.
The reasons for this large diversity lies in the simple facts that 
(i) the flux levels of the various constituents, as well as the 
sensitivity of the available detectors/telescopes are vastly different, 
and (ii) systematic or automatic searches for SNe are only made in 
the optical. Prior to the beginning of automatic searches in the 
late 1980s \cite{colgate87}, about two dozen SNe were reported per 
year in the optical. This number has now grown to about several 
hundred per year (e.g. 248 confirmed detections in the year 
2001\footnote{from the Asiago Supernova Catalogue:
http://merlino.pd.astro.it/$\sim$supern}). The detection of two dozen 
neutrinos from SN~1987A \cite{bionta87,hirata87} was only possible 
because of its close proximity at a distance of about 50 kpc in the 
Large Magellanic Cloud. SNe have been detected up to a redshift
of $z\sim1.7$ (SN 1997ff \cite{riess01}) in the optical and IR, 
and up to $z\sim0.022$ (SN 1988Z; $\sim$100 Mpc) in both the 
radio \cite{vandyk93} and X-rays \cite{fabian96}.

Based on the presence or absence of hydrogen lines in their
spectra, SNe are classified as Type II and I, respectively
(see the Chap. by Turatto and references therein).
No Type Ia SN has been detected to date in either the radio
or in X-rays. These SN are believed to be nuclear detonations of
carbon+oxygen (C+O) white dwarfs when they exceed the Chandrasekhar limit 
through accretion (\cite{nomoto84,woosley86} and references therein). 
Due to recurrent pulsations of the progenitor, fast shells 
(several $\times100~{\rm km~s}^{-1}$) are ejected from the progenitor.
During periods of quiescence, no stellar wind is blown into the
CSM due to the low mass of the progenitor. This leads to the formation 
of a complicated, yet low density CSM structure with interacting shells of 
different expansion speeds. The relatively slow expanding ejecta
($\ltap 5,000~{\rm km~s}^{-1}$) can hence not form a shock region
that could give rise to thermal X-ray or radio emission (see below). 
Since no X-ray emission has been recorded from a Type Ia SN to date,
and is neither expected at a detectable flux level, we will not discussed 
Type Ia's here further.

This leaves us still with a zoo of different types: Ib, Ic, IIP,
IIL, IIb, and IIn all of which are believed to be the result of the
core collapse of a massive star ($\gtap10~M_{\odot}$ ZAMS;
e.g. \cite{burrows95a,hashi93} and references therein).
The dividing line between these types is not always clear, and there
are numerous examples whereby an original designation was later
changed  as the spectra and the light curves evolved in time
(SN 1998bw, for example, was classified as both Ib and Ic at different 
times during its evolution).
The more detailed information is available on a particular SN, 
the more difficult it becomes to `squeeze' it into a category 
of known types. This has led to the introduction of the 
`peculiar' SNe, abbreviated `pec', and it is not so surprising that 
two of the best studied SNe, 1987A and 1993J, have both 
obtained the `pec' distinction from several authors 
(e.g. \cite{fil94,silvia90}). 

Here follows a verbatim quote from Kurt Weiler: 
{\it ``My general rule of thumb is: any well observed SN is peculiar. 
Only the poorly observed ones fit nice classifications -- for classifications, 
bad data is good. Think of it this way: at a distance, and in the dark,
all cats fit a nice category.  With close examination, and in the
light, they are all individuals.''}

The X-ray luminosities of all detected SNe are in the range
$10^{37}$--$10^{41}~{\rm ergs~s}^{-1}$. A compilation of all 15 X-ray 
SNe is given in Tab.\ 1\footnote{a frequently updated list of X-ray SNe
and references are available at: \\ 
http://xray.astro.umass.edu/supernovae\_list.html}.
They dominate the total luminosity of the
SNe starting at an age of about one year. The X-ray emission, as well
as the radio emission, are largely the result of the interaction of the
ejecta with the CSM, which is present as a
result of the wind of the SN progenitor \cite{chevfran94,suzno95}. 
Progenitors of Type II SNe following core collapse of massive massive stars
have high mass loss rates ($\dot{M} \sim 10^{-4}$--$10^{-6}$~M$_{\odot}$~yr$^{-1}$) 
and low wind velocities of typically $v_{\rm w}\sim10~{\rm km~s}^{-1}$.
Type Ib/c SNe, likely to originate from more compact stars, 
have lower mass-loss rates 
($\dot{M}\sim10^{-5}$--$10^{-7}~M_{\odot}~{\rm yr}^{-1}$) and 
significantly higher wind velocities of $v_{\rm w}\ltap1,000~{\rm km~s}^{-1}$.

Shocks are formed as the ejecta plow into the CSM: (i) the 
`circumstellar' or `forward shock' (also called `blastwave'), 
and (ii) the so-called `reverse shock'. 
In the early phase after the SNe, the speed of the 
ejecta (i.e., of the circumstellar shock) is of the order of 
$\sim10^4~{\rm km~s}^{-1}$.
The speed of the reverse shock is $\sim$10$^3~{\rm km~s}^{-1}$ lower.
Depending on the density profile of the ejecta, as they emerge from
the star, the density {\it behind} the reverse shock (i.e.\ at {\it
larger} radii than that of the reverse shock) can be 5--10 times higher
than the density {\it behind} the circumstellar shock (i.e.\ at {\it
smaller} radii than the circumstellar shock front; cf. Fig.~1
in the Chap. by Chevalier \& Fransson). The temperature 
behind the circumstellar shock can be as high as 10$^{9-10}$~K,
whereas the temperature behind the reverse shock is significantly lower
(10$^{7-8}$ K). The soft X-ray ($\ltap 5$~keV) emission is therefore
generally explained in terms of thermal radiation from the reverse
shock, whereas the higher energy X-rays ($\gtap10$~keV) are
more likely to arise in the forward shock region.  Because of the
much higher density in the reverse shock region, the X-ray emission
from this region will dominate that of the forward shock by the time
that the expanding shell has become optically thin enough to allow the
X-rays from the reverse shock region to reach the observer.

Both radio and X-ray emission are the result of the interaction of the
ejecta with the CSM.  The radio emission is believed to come
from the region behind the circumstellar shock (see the Chap.
by Chevalier \& Fransson and references therein), where the
temperature is very high and the density low. In Fig.\ \ref{lightcurve}, 
we plot the 6~cm radio and soft X-ray band (0.1--2.4 keV)
light curves for several SNe. Note that the radio emission is
absorbed at early times (see below), but later on, the radio
and X-ray emission are roughly proportional and show similar rates 
of decline.

The radio emission, in general, turns on in a late phase (months to
years after the SNe). Chevalier \cite{chev82} proposed that the
circumstellar shock generates the relativistic electrons and enhanced
magnetic field necessary for synchrotron radio emission. The CSM, which 
is ionized by the UV produced during the outburst, absorbs this radio 
emission in the early phase. However, as the circumstellar shock moves
outward, progressively less ionized material is left between the
circumstellar shock and the observer, and absorption of the radio
emission decreases. The observed radio flux density rises accordingly,
first at high frequencies and subsequently (weeks to months later) at
lower frequencies (cf, e.g.\ Fig.\ 2 in the Chap. by Sramek \& Weiler).  
At the same time, radio emission from the circumstellar shock region is
decreasing slowly as the circumstellar shock expands. When radio
absorption has become negligible, the radio light curve reflects this
decline.

The signatures of circumstellar interaction in the radio, optical, and
X-ray regimes have been found for a number of Type II SNe, such as 
the Type II-L SNe~1979C \cite{weiler86,weiler91,fesmat93,immler98a} 
and 1980K \cite{canizares82,weiler86,weiler92,fesbeck90,leib91}. 
In the case of IIn SNe, H$\alpha$ typically exhibits a very
narrow component (FWHM $\ltap200~{\rm km~s}^{-1}$) superimposed on a base of
intermediate width (FWHM $\sim1,000$--$2,000~{\rm km~s}^{-1}$).
Sometimes a very broad component 
(FWHM $\sim5,000$--$10,000~{\rm km~s}^{-1}$) is also present
\cite{fil97}. The narrow optical lines are clear evidence for
the presence of slowly moving dense circumstellar matter, probably
photo-ionized by the intense flash of UV radiation.  Examples of Type
IIn SNe are SN~1986J \cite{rupen87,weiler90,leib91,bregpil92,houck98}, 
SN~1995N \cite{ben95,lew96,vandyk96}, and SN~1998S 
\cite{filmor98,gar98,li98,vandyk99,po01a}. 
Each of these SNe are discussed in more detail below.

Several nearby Type Ib/c SNe have been detected in the radio: SN~1990B
\cite{vandyk93}, SN~1994I \cite{vandyk94}, SN~1997X \cite{vandyk01},
SN~1998bw \cite{galama98,kulkarni98} and SN~2002ap \cite{berger02}.
SN~1994I \cite{immler98b,immler02a}, SN~1998bw \cite{pian99,pian00}
and SN~2002ap \cite{pascual02} have also been detected in X-rays. 
These SNe may be special cases --- in particular, the strong 
evidence for association between SN~1998bw and GRB~980425 makes it 
unique (see the Chaps. by Galama and Frontera).

\newpage
\section{X-Ray Production Mechanism}
\label{theory}

As discussed above, the most important process for the generation
of a substantial amount of X-ray emission is the interaction of the
ejecta with the CSM. However, depending on the age of the SNe, there
can be other mechanism for the production of X-ray emission: 
(i) radio activity of the ejecta,
(ii) an initial burst of X-rays during the break-out of the
shock from the surface of the star, producing short time-scale
($\sim$1,000~sec) hard X-rays ($\sim100~{\rm keV}$) and
a very soft ($\sim0.02~{\rm keV}$) black-body continuum,
(iii) inverse Compton scattering of relativistic electrons
with UV photons produced by the outburst, and
(iv) pulsar-driven X-ray emission from the SN remnant in the case
of a hot, fast spinning neutron star (if the latter is formed 
in the process of core collapse). This model is applicable to
older SNRs, such as the Crab nebula.

Cases (ii)--(iv) will not be further discussed since they 
are not relevant to the X-ray SNe reviewed in this chapter,
and we will restrict ourselves to give a brief overview
of the CSM interaction and radioactive decay.

\subsection{The Circumstellar Interaction}

The thermal X-ray luminosity, $L_{\rm x}$, produced by the shock heated 
CSM is the product of the emission measure, EM, and the cooling function, 
$\Lambda(T,Z,\Delta E)$, where $T$ is the CSM plasma temperature,
$Z$ represents the elemental abundance distribution, and $\Delta E$ is 
the X-ray energy bandwidth. For spherically symmetric conditions 
$L_{\rm x} = \Lambda(T,Z,\Delta E) {\rm d}V n^2$, where d$V$ is the volume, 
$n = \rho_{\rm csm}/m$ is the number density of the shocked CSM and $m$ 
is the mean mass per particle ($2.1\times10^{-27}$~kg for a H+He plasma). 
Assuming a constant supernova shock speed $v_{\rm s}$, 
$r=v_{\rm s}t$, where $t$ represents the time elapsed since the explosion. 
If the CSM density $\rho_{\rm csm}$ is dominated by a wind blown 
by the progenitor star of the supernova, the continuity equation requires 
$\dot{M} = 4\pi r^2 \rho_{\rm w}(r) \times v_{\rm w}(r)$ through a sphere 
of radius $r$. After the SN shock plows through the CSM, its density is 
$\rho_{\rm csm} = 4\rho_{\rm w}$ \cite{flc96}. We thus obtain 
$L_{\rm x} = 4/(\pi m^2) \Lambda(T) \times (\dot{M}/v_{\rm w})^2\times(v_{\rm s} t)^{-1}$.
We can hence use the observed X-ray luminosity at time t after the outburst 
to measure the ratio $\dot{M}/v_{\rm w}$.

Each X-ray measurement at time $t$ is related to the corresponding 
distance $r$ from the site of the explosion. This site had been reached 
by the wind at a time depending on $v_{\rm w}$, or the age of the wind 
$t_{\rm w} = t v_{\rm s}/v_{\rm w}$. 
Usually, $v_{\rm s} \gg v_{\rm w}$ so that with $t$ only a few years we 
can look back quite a large time span in the evolution of the progenitor's 
wind and can use our measurements as a `time machine' to probe the 
progenitor's history. Assuming that $v_{\rm w}$ did not change over 
$t_{\rm w}$, we can even directly measure the mass loss rate back in time.
Integration of the mass-loss rate along the path of the expanding shell gives 
the mean density inside a sphere of radius $r$. For a constant wind velocity 
$v_{\rm w}$ and mass-loss rate $\dot{M}$, a 
$\rho_{\rm csm} = \rho_0 (r/r_0)^{-s}$ profile with $s=2$ is expected.

After the expanding shell has become optically thin, it is expected that 
emission from the SN ejecta itself, heated by the reverse shock, dominates 
the X-ray output of the interaction regions due to its higher emission
measure and higher density. For uniformly expanding ejecta
the density structure is a function of its expansion velocity,
$v$, and the time after the explosion, $t$: 
$\rho_{\rm sn}  = \rho_0 (t/t_0)^{-3} (v/v_0)^{-n}$ with $\rho_0$ 
the ejecta density at time $t_0$ and velocity $v_0$ \cite{flc96}. 
For a red supergiant progenitor, the power-law is rather steep with 
index $n\sim20$ \cite{flc96,suzno95}. For constant $n$, the radius of 
the discontinuity surface between the forward and the reverse shock 
evolves in time $t$ with $r_{\rm c} \propto t^m$, where $m=(n-3)/(n-s)$
is the deceleration parameter.

\subsection{X-Ray Emission from Radioactive Decay}

The SN ejecta are radioactive. Gamma-rays due to the decay of Ni$^{56}$ 
(half live 6.1 days) $\rightarrow$ Co$^{56}$ (half live 77 days) 
$\rightarrow$ Fe$^{56}$ and Ni$^{57}$ (half live 36 hours) $\rightarrow$
Co$^{57}$ (half live 270 days) $\rightarrow$ Fe$^{57}$ could, in principle, 
be detectable during the first few years after the SNe. The dominant
lines produced in the decay of Co$^{56}$ are 511 keV (40\%), 847 keV
(100\%), 1.04 MeV (15\%) 1.24 MeV (66\%), 1.76 MeV (15\%), 2.02 MeV
(11\%), 2.60 MeV (17\%), and 3.25 MeV (13\%). The dominant lines
produced in the decay of Co$^{57}$ are 14 keV (9\%), 122 keV (87\%),
and 136 keV (11\%).  The numbers in brackets represent the probablities
that for each decaying nucleus the $\gamma$-ray in question is
produced.

Several of these lines have been detected from SN~1987A (see the 
Chap. by McCray and references therein) but not from any other SN. 
This, of course, is not surprising given the very modest flux levels 
of the $\gamma$-rays and the large distances to all SNe.  
Hard ($\gtap10~{\rm keV}$) X-rays were observed from SN~1987A as 
early as 168--191 days after the outburst \cite{sun87,dotani87}. 
At least part of these X-rays are believed to be the result of 
Compton scattering of $\gamma$-rays \cite{kumagai88}. The light curves and
spectral evolution of the X-rays and $\gamma$-rays are reasonably well
understood (up to $\sim300$ days after the outburst) if one assumes
mixing of Co$^{56}$ into the hydrogen rich envelope. The X-ray light
curve at later times may be more problematic \cite{kumagai89}.
Whether a neutron star or a black hole was formed during the outburst 
of SN~1987A is still a matter of debate.

In case that a Type Ia SNe occurred within a distance of $\sim$50 kpc,
one would also expect to see $\gamma$-rays from the radioactive
by-products of the detonation (subsonical speed of the shock through
the progenitor) or deflagration (supersonical speed) of the C+O white 
dwarf. This, however, has not been observed as yet.

\begin{table}[p!]
\caption{List of X-Ray Supernovae}
\scriptsize{
\begin{center}
\renewcommand{\arraystretch}{1.4}
\setlength\tabcolsep{5pt}
\begin{tabular}{llcccccrr}
\hline\noalign{\smallskip}
No. & Supernova & Type & $d$  & $t-t_0^{\star}$ & \fx$^{\dag}$ 
 & \Lx$^{\ddag}$ & Instrument & Ref. \\
 & Galaxy & & [Mpc] & [days] & [$10^{-14}$] & [$10^{38}$] \\
\noalign{\smallskip}
\hline
\noalign{\smallskip}
\phantom{0}1. & {\bf SN 1978K}  & II & 4.5 & 4,500 & 204   &  48      
	& \R, \A & \cite{petre94,schlegel95,schlegel96,schlegel99a} \\
 & NGC 1313  & & & & \multicolumn{2}{c}{(0.1--2.4~keV)} & \X \\ 
\noalign{\smallskip}
\phantom{0}2. & {\bf SN 1979C}  & IIL   & 17.1 & 5,900 &  3.9  &  14      
	& \R, \C  & \cite{immler98b,kaaret01,ray01} \\
 & NGC 4321 & & & & \multicolumn{2}{c}{(0.1--2.4~keV)} & \X \\ 
\noalign{\smallskip}
\phantom{0}3. & {\bf SN 1980K}  & IIL   & 5.1 &     35 &  1.5  &  0.5     
	& \E, \R & \cite{canizares82,schlegel94,schlegel95} \\
 & NGC 6946 & & & &\multicolumn{2}{c}{(0.1--2.4~keV)} \\ 
\noalign{\smallskip}
\phantom{0}4. & {\bf SN 1986J}  & IIn & 9.6 & 3,300 &  119  &  140     
	& \R, \A & \cite{bregpil92,chugai93,houck98,schlegel95} \\
 & NGC 891  & & & &\multicolumn{2}{c}{(0.1--2.4~keV)} \\ 
\noalign{\smallskip}
\phantom{0}5. & {\bf SN 1987A}  & IIP   & 0.05 & -- & -- & --
	& \R, \C & 	\cite{burrows00,dennerl01,gore94,hasinger96} \\
 & LMC & & & & & & \X & $^{\S}$ \\
\noalign{\smallskip}
\phantom{0}6. & {\bf SN 1988Z}  & IIn & 89   & 2,370 &  0.9  &  110     
	& \R & \cite{aretxaga99,fabian96} \\
 & \multicolumn{2}{l}{MCG+03-28-022}& & & \multicolumn{2}{c}{(0.1--2.4~keV)} \\ 
\noalign{\smallskip}
\phantom{0}7. & {\bf SN 1993J}  & IIb   & 3.6 &      6 &  130  &  20      
	& \R, \A & \cite{immler01,suzno95,zimmermann94} \\
 & NGC 3031 & & & & \multicolumn{2}{c}{(0.5--2~keV)} \\ 
\noalign{\smallskip}
\phantom{0}8. & {\bf SN 1994I}  & Ic    & 7.7 &     79 &  2.3  &  1.6     
	& \R, \C & \cite{immler98b,immler02a} \\
 & NGC 5194 & & & &\multicolumn{2}{c}{(0.3--2~keV)} \\ 
\noalign{\smallskip}
\phantom{0}9. & {\bf SN 1994W}  & IIP   & 25   & 1,180 &  11   &  85      
	& \R & \cite{schlegel99b} \\
 & NGC 4041 & & & & \multicolumn{2}{c}{(0.1--2.4~keV)} \\ 
\noalign{\smallskip}
10. & {\bf SN 1995N}  & IIn & 24   &    440 &  40   &  175     
	& \R, \A & \cite{fox00} \\
 & \multicolumn{2}{l}{MCG-2-38-017}& & & \multicolumn{2}{c}{(0.1--2.4~keV)} \\ 
\noalign{\smallskip}
11. & {\bf SN 1998bw} & Ic    & 38   &    0.4 & 23--40 & 40--70  
	& \B & \cite{pian99,pian00} \\
 & \multicolumn{2}{l}{ESO 184-G82} & & & \multicolumn{2}{c}{(2--10~keV)} \\ 
\noalign{\smallskip}
12. & {\bf SN 1998S}  & IIn   & 17   &     64 &  27   &  92      
	& \C & \cite{po01a} \\ 
 & NGC 3877 & & & & \multicolumn{2}{c}{(2--10~keV)} & \X \\ 
\noalign{\smallskip}
13. & {\bf SN 1999em} & IIP   & 7.8  &      4 &  1.0   &  0.7     
	& \C & \cite{po01a} \\ 
 & NGC 1637 & & & & \multicolumn{2}{c}{(2--10~keV)} \\ 
\noalign{\smallskip}
14. & {\bf SN 1999gi} & IIP   & 8.7  &     29 &  0.1  &  0.1     
	& \C & \cite{schlegel01} \\ 
 & NGC 3184 & & & & \multicolumn{2}{c}{(0.5--10~keV)} \\
\noalign{\smallskip}
15. & {\bf SN 2002ap} & Ic   & 10  & 4 &  0.2  &  0.2
	& \X & \cite{pascual02} \\ 
 & NGC 628 & & & & \multicolumn{2}{c}{(0.1--10~keV)} \\
\noalign{\smallskip}
\hline
\end{tabular}
\begin{list}{}{}
\item $^{\star}\phantom{0}$ day after outburst corresponding 
to the maximum observed X-ray flux
\item $^{\dag}\phantom{0}$ maximum observed X-ray flux
in units of $10^{-14}~{\rm ergs~cm^{-2}~s^{-1}}$ 
\item $^{\ddag}\phantom{0}$ maximum observed X-ray luminosity 
in units of $10^{38}~{\rm ergs~s^{-1}}$ 
\item $^{\S}\phantom{0}$ X-ray flux still rising
\end{list}
\label{sne_list}
\end{center}
}
\end{table}

\begin{figure}[h!]
\begin{center}
\includegraphics[width=0.999\textwidth]{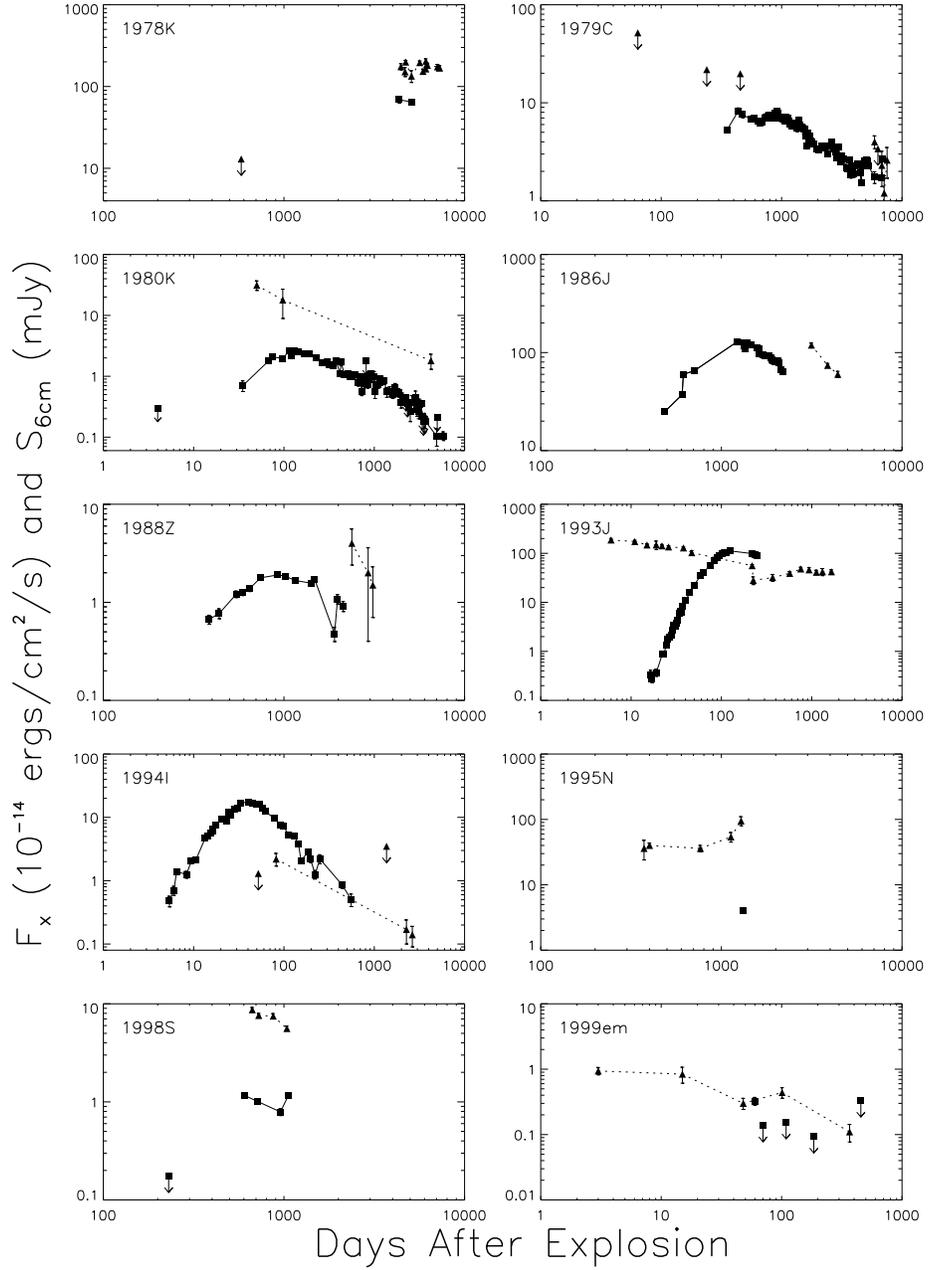}
\end{center}
\caption[]{6~cm radio (\V\ and {\sl ATCA}; filled boxes, solid lines) 
and soft X-ray band lightcurves ($0.2$--$2.4$~keV; triangles and dashed lines) 
of nine Type II and one Ic SNe.}
\label{lightcurve}
\end{figure}

\newpage
\section{Overview of X-ray Supernovae}
In the following section we will give a brief overview 
of X-ray observations of all X-ray SNe except SN~1987A,
which is discussed in detail in the Chap. by McCray.

\subsection{SN~1978K in NGC~1313}
\label{1978k}
Due to its discovery in the early phase of the \R\ mission and 
the relatively high flux level, SN~1978K is one of the best studied 
SN in the X-ray regime.
While an \E\ observation $\sim1.5$ years after the outburst 
only provided an upper limit to the soft (0.5--2~keV) band X-ray
luminosity ($L_{\rm x}\ltap2\times10^{38}~{\rm ergs~s}^{-1}$ 
\cite{schlegel96}), SN~1978K has been successfully monitored 
with \R\ on 13 different dates ranging from $\sim12$ to $20$ years 
after the outburst \cite{schlegel96,schlegel99a}. During this 
period, the SN showed no apparent evolution in X-ray flux
($L_{\rm x}\sim4\times10^{39}~{\rm ergs~s}^{-1}$).
\A\ GIS/SIS and \R\ PSPC observations
provided the first broad-band (0.1--10~keV), medium resolution
X-ray spectra of a young SN \cite{petre94}. The spectra showed no 
emission lines and could be equally well described by either a 
thermal bremsstrahlung spectrum or thermal plasma emission with a 
temperature of $\sim3$~keV and sub-solar abundances ($Z\sim0.2Z_{\odot}$). 
Alternatively, the spectrum could be described by a power with 
photon index $\Gamma\sim2.2$ and an absorbing column of 
$N_{\rm H}\sim1\times10^{21}~{\rm cm}^{-2}$.
From the X-ray lightcurve the index $n$ for the ejecta power-law
distribution $\rho_{\rm ejecta} \propto r^{-n}$ could be constrained 
to be in the range 4--12 for an assumed CSM density gradient
$\rho_{\rm csm} \propto r^{-s}$ with index $s=1.5$--$2$
\cite{schlegel99a}. Estimates for the mass-loss rate are
$\dot{M}\sim1\times10^{-4}~M_{\odot}~{\rm yr}^{-1}$, 
as is expected for a massive progenitor.
A recent \X\ observation showed that SN~1978K is still at the
same flux level as observed during the last $\sim22$ years
and shows no apparent evolution
($L_{\rm x}=4\times10^{39}~{\rm ergs~s}^{-1}$ \cite{immler02b}).
The superior photon collecting area and spectral resolution 
of the \X\ instruments with a total of $\sim20,000$ net counts 
from SN~1978K in the combined EPIC-pn and MOS spectra demonstrates
the dramatic technological progress that has been made in
the development of X-ray instruments over the last two decades. 
The best-fit model to the spectrum gave a two-temperature thermal
emission component with $kT_{\rm low}\sim0.8$~keV and 
$kT_{\rm high}\sim3$~keV. The \X\ data clearly show, for the 
first time, emission from the forward (high temperature component) 
and reverse shock region (low temperature component).

\subsection{SN~1979C in NGC~4321 (M100)}
\label{1979c}
\begin{figure}[t!]
\begin{center}
\includegraphics[width=.7\textwidth]{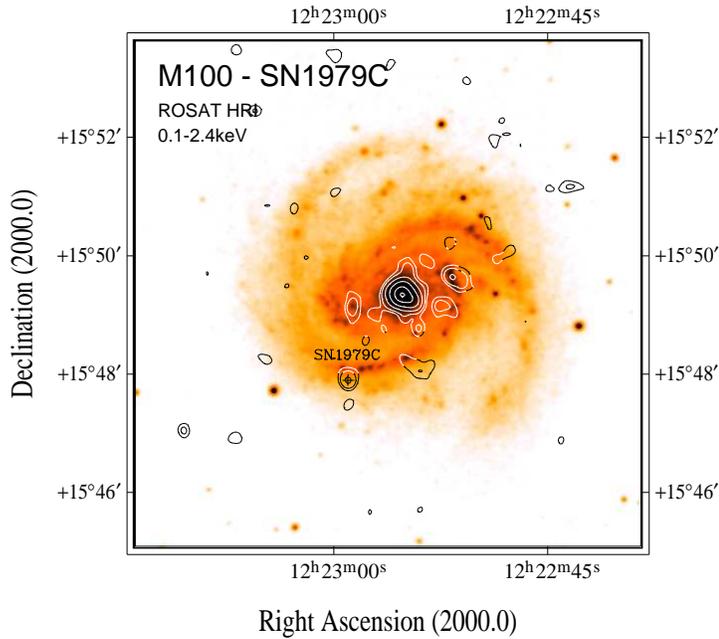}
\end{center}
\caption{\R\ HRI soft (0.1--2.4~keV) band X-ray contours of SN~1979C in M100,
overlayed onto an optical image.}
\label{1979c_hri}
\end{figure}

An X-ray source was discovered in a \R\ HRI observation of
M100 at the position of SN~1979C $\sim16$ years after the outburst
(see Fig.\ \ref{1979c_hri}), with a (0.1--2.4~keV) luminosity of 
$L_{\rm x}=1.3\times10^{39}~{\rm ergs~s}^{-1}$ 
\cite{immler98a}, and in a \R\ HRI follow-up observation
$\sim19$ years after the outburst.
For three earlier \E\ observations, taken on days 64, 239 
and 454 after the outburst, only $3\sigma$ upper limits
could be established
($1.8 \times 10^{40}~{\rm ergs~s}^{-1}$, 
$7.6 \times 10^{39}~{\rm ergs~s}^{-1}$ and 
$6.9 \times 10^{39}~{\rm ergs~s}^{-1}$, respectively). 
The \R\ data imply a mass-loss rate of 
$\dot{M}\sim1\times10^{-4}~M_{\odot}~{\rm yr}^{-1}$,
similar to mass-loss rates of other massive SN progenitors
(e.g. SNe 1978K, 1986J, 1988Z and 1998S)
and in agreement with the mass-loss rate inferred from \V\
radio observations \cite{weiler91}.
Since neither a \R\ PSPC `last light' observation (requested by the 
author), nor \A\ and \C\ pointings \cite{kaaret01,ray01} gave enough photon 
statistics to constrain the spectral properties, no information is 
available about the spectral characteristics of the X-ray emission. 
Both the \R\ PSPC and \C\ observations, however, showed that the
emission is rather soft ($<2$~keV) and likely originates from
shock-heated material in the reverse shock.
The combined \R, \A\ and \C\ data indicate a rather slow X-ray rate 
of decline (decrease from $L_{\rm x}=1.3\times10^{39}~{\rm ergs~s}^{-1}$
to $0.9\times10^{39}~{\rm ergs~s}^{-1}$ over 4.3 years in the 0.1--2.4~keV band
\cite{kaaret01}) consistent with the rate of decline ($t^{-0.7}$) observed 
in the radio regime over the first $\sim10$ years after the outburst \cite{weiler91}.
Since both the X-ray and radio emission processes are considered 
to be linked to the interaction of the SN shock with the 
high-density envelope of matter (see Sec.~1), one would expected 
the X-ray evolution to follow the observed radio evolution.
\vfill

\subsection{SN~1980K in NGC~6946}
\label{1980k}
Historically, the IIL SN~1980K in NGC~6946 was the first SN 
discovered in X-rays \cite{canizares82}. SN~1980K was detected
in an \E\ IPC observation with a (0.2--4~keV) luminosity of
$L_{\rm x}=4.8\times10^{38}~{\rm ergs~s}^{-1}$ 35 days after
the outburst \cite{canizares82,schlegel95}. A follow-up
observation on day 82 showed that the X-ray source has
faded by a factor of $\sim5$. Despite the rather large 
IPC error box of $\sim1'$, the positional coincidence
of the X-ray source with the optical position of the SN was
later confirmed with the \R\ PSPC \cite{schlegel94}.
Spectral analysis of the \E\ and \R\ data showed that the
emission was rather soft (\E: $kT\sim0.5$~keV \cite{canizares82}; 
\R: $kT\sim0.4$~keV \cite{schlegel95}) 
and in relatively good agreement with theoretical predictions
($kT\sim0.7$~keV \cite{chevfran94}). Given the low photon
statistics, physical processes as to the production
of the X-rays could not be constrained. The data, however,
could be successfully used to give estimates as to the mass
loss rate of the SN progenitor 
($\dot{M}\sim0.5$--$5\times10^{-6}~M_{\odot}~{\rm yr}^{-1}$) 
and to demonstrate that SNe do not significantly contribute to 
the overall diffuse X-ray background.

\subsection{SN~1986J in NGC~891}
\label{1986j}
SN~1986J was observed with the \R\ HRI and with \A\
over a period covering $\sim9$--$13$ years after the outburst
\cite{bregpil92,chugai93,houck98}. Contrary to many other
X-ray SNe (e.g. SNe 1978K, 1993J, 1994I and 1998S), 
SN~1986J showed a rather fast X-ray rate of
decline ($L_{\rm x} \propto t^{-2}$ \cite{houck98}).
Two \A\ spectra indicated that the emission was rather
hard ($kT=5$--$7$~keV) compared to other X-ray SNe
(cf. Sec.~\ref{1980k} and \ref{1979c}). The \A\ spectra 
also clearly showed a Fe-K emission line at $6.7$~keV with 
a width of $<20,000~{\rm km~s}^{-1}$ (FWHM).
The spectral properties and rate of decline were used
to test two different models that could account
for the observed X-ray emission: in the CSM interaction
model the X-ray emission is thought to originate in the
reverse shock running through the outer layers of the
SN ejecta \cite{chevfran94}. While both the predicted
X-ray luminosity ($\sim10^{40}~{\rm ergs~s}^{-1}$), 
mass-loss rate ($\sim10^{-4}~M_{\odot}~{\rm yr}^{-1}$ 
for an assumed wind velocity of $10~{\rm km~s}^{-1}$) 
and line width for the shocked plasma 
($\sim 10,000~{\rm km~s}^{-1}$ FWHM) are in agreement
with the X-ray data, the CSM model predicts a temperature
significantly lower ($\sim1$~keV) than observed.
An alternative model proposed by Chugai \cite{chugai93}
assumes that the emission arises from shocked interstellar
clumps after the forward shock plows through the CSM.
This model predicts a higher temperature of the shocked 
regions which are reconciled by the data, and a similar 
mass-loss rate for the progenitor. Potential problems, 
however, arise in the fact that the shocked clouds should
have narrow emission lines (some $\times100~{\rm km~s}^{-1}$).
While narrow emission lines have in fact been observed
in optical spectra of SN~1986J (e.g. H$\alpha$ emission
lines with a FWHM of $530~{\rm km~s}^{-1}$ 
\cite{rupen87,leib91}), questions arise as to how
the clouds can account for both the optical and X-ray emission,
which require a shock velocity of approx. $2,000~{\rm km~s}^{-1}$ for
the observed temperature \cite{houck98}.

\subsection{SN~1988Z in MCG+03-28-011}
\label{1988z}
At a distance of $\sim89$~Mpc (assuming 
$H_0=75~{\rm km~s}^{-1}~{\rm Mpc}^{-1}$), SN~1998Z is the most 
distant and most luminous X-ray 
($L_{\rm x}\sim10^{41}~{\rm ergs~s}^{-1}$ 6.5~years after the
outburst) and radio SN 
($L_{\rm 6cm}\sim9\times10^{32}~{\rm ergs~s}^{-1}~{\rm Hz}^{-1}$ 
on day 1,253) discovered to date \cite{fabian96,vandyk93}.
Given the observed X-ray luminosity, a CSM density of 
$\sim10^6~{\rm cm}^{-3}$ was inferred at a radius of 
$r=5\times10^{16}$~cm from the site of the explosion 
(corresponding to the date of the X-ray observation) 
\cite{fabian96}.
The radio data further implied that the dense cocoon 
resulted from a high mass-loss rate of the progenitor of 
$\dot{M}\sim10^{-4}~M_{\odot}~{\rm yr}^{-1}$ in the late stage 
of the massive ($20$--$30~M_{\odot}$) progenitor \cite{vandyk93}.
Two \R\ HRI follow-up observations \cite{aretxaga99} showed that the SN 
rapidly faded by a factor of $\sim5$ within 2 years (see Fig.\ \ref{lightcurve}).

\subsection{SN~1993J in NGC~3031 (M81)}
\label{1993j}
\R\ observations of SN~1993J gave strong observational support
for models that attribute the X-ray emission to the interaction
of an outgoing and reverse shock with the CSM.
Whereas the early X-ray spectrum on day 6 was hard ($T\sim10^{8.5}$~K)
and absorbed by the Galactic foreground column only
($N_{\rm H}\sim4\times10^{20}~{\rm cm}^{-2}$),
a softer component ($T\sim10^{7}$~K) dominated at day $\sim$200
\cite{zimmermann94,flc96}. \A\ data from day 8 in the broader 
1--10~keV X-ray band were also best characterized by
hard ($\sim10^{8}$~K) thermal emission \cite{tanaka93}.
Very hard X-ray photons were recorded during days 9--15
and 23--36 using OSSE onboard the {\sl Compton Gamma-Ray Observatory}.
During these dates, SN~1993J reached (50--150~keV band) 
luminosities of $L_{\rm x}=5.5\times10^{40}~{\rm ergs~s}^{-1}$ and 
$3.0\times10^{40}~{\rm ergs~s}^{-1}$, respectively, before falling
below the OSSE detection threshold during a longer observation on 
days 93--121 \cite{leising94}.
The harder component is due to the circumstellar shock region, whereas 
the softer component must have come from the region behind the reverse 
shock. The emergence of this radiation could be attributed to the 
decreased absorption by a cool shell \cite{flc96}.
The early ($<200$ days) \R\ and \A\ data were successfully
used to estimate the mass-loss rate of the SN~1993J progenitor
($\dot{M} \sim 4\times10^{-5}~{\rm M}_{\odot}~{\rm yr}^{-1}$) 
and indicate that the CSM density
profile might be flatter ($\rho_{\rm csm} \propto r^{-s}$ with
$1.5 \ltap s \ltap 1.7$) than expected for spherically symmetric
conditions ($s=2$) \cite{flc96}.

Intriguing new results came form the analysis of the
entire \R\ data set, covering a period from 6 days to 7 
years after the outburst of SN~1993J. The combined \R\
PSPC and HRI lightcurve is best fitted by a 
$L_{\rm x} \propto t^{-0.27}$ X-ray rate of decline
(cf. Fig.\ \ref{lightcurve}). Since each X-ray 
observation at time $t$ is related to the corresponding 
distance $r$ from the site of the explosion, 
$r=v_{\rm s} \times t$ (with shock front velocity $v_{\rm s}$),
and to the age of the stellar wind, 
$t_{\rm w} = t v_{\rm s}/v_{\rm w}$, the \R\ measurements
could be used as a `time machine' to look back into the
history of the stellar wind. 
During the observed period, the SN shell has reached a radius 
of $3\times10^{17}$~cm from the site of the explosion, corresponding 
to $\sim10^4$ years in the progenitors stellar wind history. 
Contrary to an expected CSM density profile of 
$\rho_{\rm csm} = \rho_0 (r/r_0)^{-s}$ profile with index $s=2$
for a constant wind velocity $v_{\rm w}$ and mass-loss rate $\dot{M}$, 
the data revealed a significantly flatter profile of 
$\rho_{\rm csm} \propto r^{-1.63}$.
After ruling out alternative scenario that might explain the data
(e.g. variations in velocity and/or temperature of the shocked CSM 
and a non-spherical geometry caused by a binary evolution of the 
progenitor) it was concluded that the mass-loss rate of the progenitor 
has decreased constantly from $\dot{M} = 4\times10^{-4}$ to 
$4\times10^{-5}~{\rm M}_{\odot}~{\rm yr}^{-1} (v_{\rm w}/10~{\rm km~s}^{-1})$.
The observed evolution clearly reflects either a decrease in the 
mass-loss rate, an increase in the wind speed or a combination of both,
indicating that the progenitor likely was making a transition from the red 
to the blue supergiant phase during the late stage of its evolution.
This scenario for the evolution of the SN~1993J progenitor has
interesting similarities with that of SN~1987A, whose progenitor 
(aka SK-69\,202) completely entered the blue supergiant phase after 
significant mass-transfer to a companion some $\times10^4$ years
prior to the explosion (see \cite{fran89,pod93} and the Chap. by McCray).

\begin{figure}[t!]
\begin{center}
\includegraphics[width=.55\textwidth]{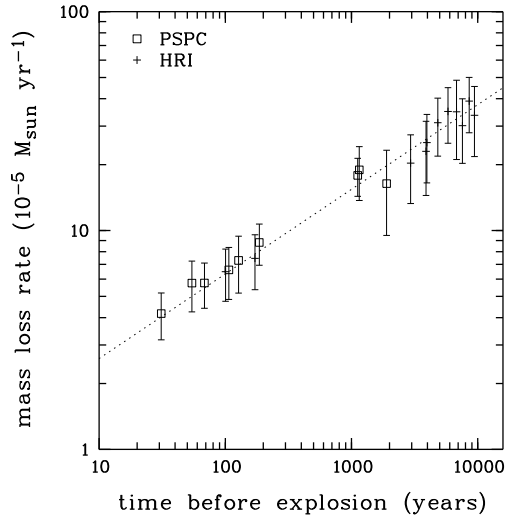}
\end{center}
\caption{Mass-loss rate history of the SN~1993J progenitor.
\R\ PSPC data are marked by boxes, HRI data are indicated by 
crosses \cite{immler01}.}
\label{1993_csm}
\end{figure}

\subsection{SN~1994I in NGC~5194 (M51)}
\label{1994I}
\begin{figure}[t!]
\begin{center}
\includegraphics[width=.75\textwidth]{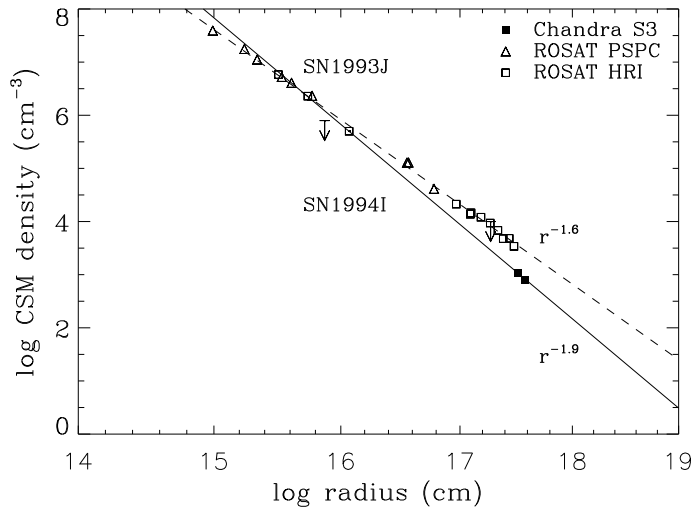}
\end{center}
\caption{CSM density profiles of SNe 1994I and 1993J as a 
function of shell expansion radius \cite{immler02a}.}
\label{1994I_1993J_rho}
\end{figure}

SN~1994I represents a unique case since it
is the first Ic SN which has been detected in soft X-rays.
Based on early \R\ HRI observations, evidence for soft (0.1--2.4~keV)
X-ray emission was found on day 83 after the outburst \cite{immler98b}.
However, due to the spatial resolution of the HRI ($5''$ on-axis)
and the close location to the X-ray bright nucleus of M51
($18''$ offset), as well
as the high level of diffuse X-ray emission in the bulge of M51,
the results were not entirely conclusive. Two follow-up observations
6--7 years after the outburst, using the superior capabilities
of \C\ in terms of spatial resolution ($0\farcs5$ on-axis)
and sensitivity revealed an X-ray source consistent with
the radio position of the SN ($0\farcs5$ offset \cite{immler02a}).
SN~1994I was detected with \C\ at a $\sim6\sigma$ level with
a luminosity of
$L_{\rm x}\sim1\times10^{37}~{\rm ergs~s}^{-1}$ in the 
0.3--2~keV band. Combined with the earlier \R\ data, the
X-ray lightcurve was parameterized as
$L_{\rm x} \propto (t-t_0)^{-s} \times e^{-\tau}$ with
$\tau\propto(t-t_0)^{-\beta}$, where the external absorption
of the emission at early times is represented by the
$e^{-\tau}$ term (`optical depth') with a subsequent
exponential rate of decline with index $s$.
A best-fit X-ray rate of decline of $L_{\rm x} \propto t^{-1.4}$
was found with index $\beta=1$ for the optical depth.
The index $s$ is in good agreement with the radio rate of
decline of SN~1994I ($s=1.6$). Interestingly, similarly fast
rates of decline have been inferred for other Ib/c SNe as well
(e.g. $s=1.4$ for SN~1998bw inferred from the tentative
hard X-ray band \B\ detection; $s=1.3$, $1.6$ and $1.5$, 
respectively, for the radio SNe~1990B 1983N and 1984L).
Assuming that the X-ray emission is due to the shocked
CSM, a mass-loss rate of
$\dot{M}\sim1\times10^{-5}~M_{\odot}~{\rm yr}^{-1}~(v_{\rm w}/10~{\rm km}^{-1})$ 
consistent with all \R\ 
and \C\ detections on days 82, 2,271 and 2,639 and a \R\ upper 
limit on day 1,368 after the outburst was derived.
Furthermore, the CSM density profile was constructed for the 
different radii corresponding to the dates of the observations. 
The combined \R\ and \C\ data gave a best-fit CSM profile of
$\rho_{\rm csm} \propto r^{-(1.9\pm0.1)}$ for $r=0.1$--$3.8\times10^{17}$~cm,
consistent with what is expected for a constant stellar wind
velocity and constant mass-loss rate of the progenitor ($r^{-2}$).
A comparison of the CSM profiles of SNe 1994I and 1993J, the
only SNe for which the CSM profile was ever constructed \cite{immler01},
is illustrated in Fig.\ \ref{1994I_1993J_rho}.

\subsection{SN~1994W in NGC~4041}
\label{1994w}
A \R\ HRI observation of SN~1994W was performed 1,180 days
after the outburst \cite{schlegel99b}. 
An X-ray source, with a luminosity of
$L_{\rm x}\sim8\times10^{39}~{\rm ergs~s}^{-1}$, was found
to coincide with the position of SN~1994W (offset $1\farcs4$).
Given the positional coincidence, the low probability of
$\sim3\times10^{-3}$ of chance coincidence with a background
or foreground object, and the indications of CSM interaction 
based on the narrow absorption lines in the optical spectra
made a detection of SN~1994W in X-rays likely.
However, since there are no other X-ray observations of the
host galaxy NGC~4041, confirmation of the detection is
still pending. Also, since SN~1994W is the only X-ray SN
which has not been detected in the radio, no additional
information about the CSM interaction is available. Interestingly, 
a CSM number density of $\rho_{\rm csm}\gtap10^8$~cm$^{-3}$ 
was inferred from the optically thick Fe{\sc ii} lines \cite{sol98}
at radii of $8.5\times10^{14}$~cm and $1.1\times10^{15}$~cm
(corresponding to days 31 and 57 after the outburst).
Based on the PCygni H$\alpha$ and H$\beta$ line profiles
of SN~1997ab (Type IIn), a similar CSM density of
$\sim10^{8}$~cm$^{-3}$ at $\sim10^{15}$~cm was estimated
\cite{sala98}.
The results are in remarkable agreement with the CSM number 
density of $\sim10^{7.5}$--$10^{8}$~cm$^{-3}$
for SNe 1994I and 1993J at $r\sim10^{15}$~cm
(cf. Fig.\ \ref{1994I_1993J_rho}) and appears to be a common 
CSM density for massive progenitors ($\gtap10~M_{\odot}$ ZAMS) 
at this distance from the site of the explosion.

\subsection{SN~1995N in MCG-2-38-017}
\label{1995n}
SN~1995N was observed with the \R\ HRI $\sim1$ and 2~years 
after the outburst, followed by an \A\ observation
some months later \cite{fox00}. The \A\ spectrum could be
well described by a $kT\sim10$~keV thermal bremsstrahlung
spectrum or a power-law with photon index $\sim1.7$,
both absorbed by the Galactic foreground column only
($N_{\rm H}\sim10^{21}~{\rm cm}^{-2}$).
Assuming that the spectrum has not changed during the 
observed period, SN~1995N has faded by approx. 30\%
during the two \R\ intervals and subsequently brightened by 
a factor of two (\A). Together with the high observed
luminosity ($\sim10^{41}~{\rm ergs~s}^{-1}$), this might be
indicative that the SN shock has plowed through a dense
and inhomogeneous CSM.

\subsection{SN~1998bw in ESO~184-G82}
\label{1998bw}
Hard (2--10~keV) X-ray emission was found with \B\ from the 
position of the unusual SN~1998bw which might be associated
with a $\gamma$-ray burst event (aka GRB~980425 \cite{pian99,pian00}).
The detection, however, is still tentative due to the large
error box of the \B\ Wide Field Camera (between $3'$ and $8'$,
90\% confidence limit) and the large probability of $\sim60\%$
for a random chance coincidence. Due to the likely association
of SN~1998bw with GRB~980425, we refer to the Chaps. by Galama
and Frontera for a more detailed review of the observations.

\subsection{SN~1998S in NGC~3877}
\label{1998s}
\begin{figure}[t!]
\rotatebox{-90}{\includegraphics[width=.65\textwidth]{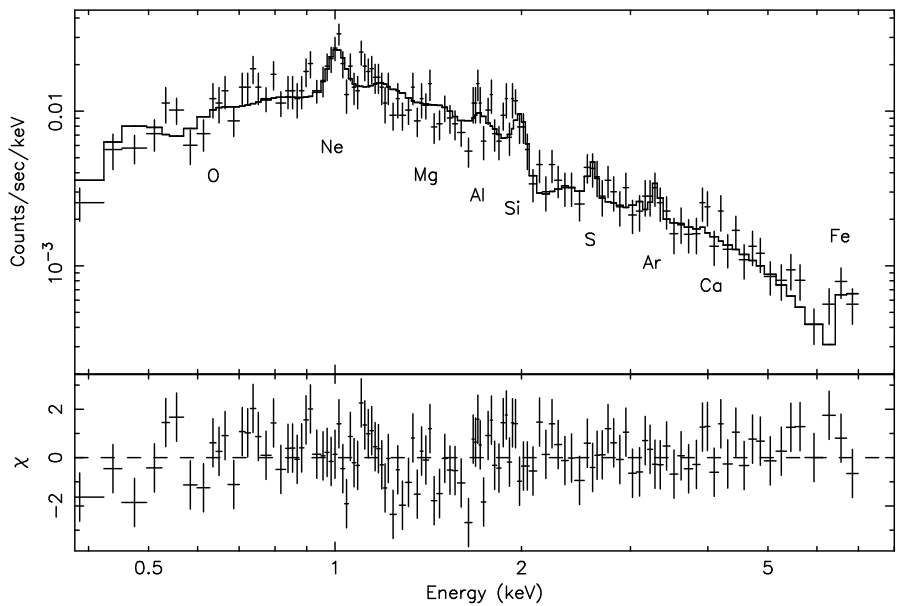}}
\caption{\C\ ACIS-S3 spectrum of SN~1998S \cite{po01a}. The solid line
gives the best fit model to the data. Labels indicate the location 
of emission lines  (see Sec.~\ref{1998s}).}
\label{1998s_spec}
\end{figure}

A wealth of X-ray data has been collected for the Type 
IIn SN~1998S with \C\ and \X\ \cite{po01a,immler02b}.
At an age of $\sim2$--$3$ years after the outburst,
SN~1998S is still bright in X-rays (\C\ data on day 668: 
$L_{\rm x} \sim 9\times 10^{39}~{\rm ergs~s}^{-1}$ \cite{po01a}; 
\X\ data on day 1,202: $L_{\rm x} \sim 3\times 10^{39}~{\rm ergs~s}^{-1}$; 
2--10~keV; \cite{immler02b}) and increasing in cm radio flux density.
The X-ray lightcurve based on five \C\ and one \X\
observation is best described by an X-ray rate of decline of
$L_{\rm x} \propto t^{-1.3}$. The inferred mass-loss rate is
$\dot{M} \sim 2 \times 10^{-4}~M_{\odot}$ yr$^{-1}$.
Spectral analysis of the \C\ data show a best-fit temperature of
$kT\sim10$~keV and high over-abundance of heavy elements, such as
Ne, Al, Si, S, Ar and Fe (3--30 over solar). The \C\ spectrum and
best-fit model are shown in Fig.\ \ref{1998s_spec}.
The high-quality spectra, in combination with theoretical modeling, 
allowed, for the first time, to use emission line ratios of elements 
produced before the core collapse and during the explosion to constrain 
the progenitor's mass. The observed O/Si ratio of 1--12, Mg/Si of 0--0.7
and Ne/Si of 0.6--14 led to an estimate of the progenitor's mass 
of $\sim18~M_{\odot}$ \cite{po01a}. 
A more recent \X\ observation revealed the emergence of a soft
($\sim0.8$~keV) component from the reverse shock in addition to
the harder component ($\sim9$~keV) already observed with \C, and 
a prominent and broad Si emission line at 1.89~keV with an equivalent 
width of $\sim50$~eV (FWHM) \cite{immler02b}.

\subsection{SN~1999em in NGC~1637}
\label{1999em}
SN~1999em is the only IIP SN detected both in the radio
and X-rays, although at a very low level
($L_{\rm 6cm}=2.2\times10^{25}~{\rm ergs~s}^{-1}~{\rm Hz}^{-1}$ 
on day 34; $L_{\rm x}=7\times10^{37}~{\rm ergs~s}^{-1}$
on day 4 \cite{po01a}). The \C\ X-ray data indicated a 
non-radiative interaction of the the SN ejecta with the CSM, 
and a low mass-loss rate of 
$\dot{M} \sim 2 \times 10^{-6}~M_{\odot}$ yr$^{-1}$.
Five \C\ observations, performed on days 4--368 after the 
outburst, showed a temperature evolution during this period, 
with a softening of the emission from originally $\sim5$~keV to 
$\sim1$~keV for an assumed thermal bremsstrahlung
spectrum. This evolution confirmed theoretical predictions
of a change in temperature and indicate a rather flat
density profile of the ejecta. 
Given an observed evolution of $L_{\rm x} \propto t^{-1}$,
the SN was below the detection limit during observations
performed on days 495 and 633 after the outburst \cite{po01a}.

\subsection{SN~1999gi in NGC~3184}
\label{1999gi}
SN~1999gi was detected in two \C\ observations on days 29 
and 54 after the outburst \cite{schlegel01}. 
Similarly to SNe 1987A and 1999em \cite{po01a}, 
SN~1999gi only reached a relatively low luminosity 
($\sim 10^{37}~{\rm ergs~s}^{-1}$; 0.5--10~keV band), 
implying a low mass-loss rate of 
$\dot{M}\sim10^{-6}~M_{\odot}$ yr$^{-1}$. Since these SNe are 
of Type IIP, it has been suggested that IIP SNe explode in a 
low-density CSM environment \cite{schlegel01}.
By contrast, progenitors of IIL (e.g. SN~1979C, 1980K), 
IIb (SN~1996J) and IIn (SN~1986J, 1988Z, 1995N, 1998S) SNe can 
deposit a substantial amount of CSM prior to the explosion.
An interesting argument was put forward by Chugai \cite{chugai97},
who described the mass-loss rate as a function of the progenitor's
zero-age main sequence star mass (ZAMS). In this scenario
the mass-loss rate rises to a peak near $\sim10~M_{\odot}$
and falls with increasing mass, leading to a IIn-IIL-IIP 
sequence. Although the X-ray and radio data collected so
far do not give sufficient statistics, they seem to
confirm this order. By collecting more data using the current 
\C\ and \X\ X-ray observatories, it remains to be seen whether 
this claim can be sustained.

\subsection{SN~2002ap in NGC~628 (M74)}
\label{2002ap}
At an age of 4 days, the Ic \cite{filchor02} SN~2002ap was detected both 
in the UV and in X-rays with the Optical Monitor 
(wavelength range 245--320~nm) and EPIC-pn onboard \X\ \cite{pascual02}.
Preliminary analysis of the EPIC-pn image showed a $3.5\sigma$
excess at a distance of $4''$ from the optical position of SN~2002ap,
below the positional error ($<10''$) at this stage of the data analysis.
Located at a distance of 10~Mpc in the host galaxy NGC~628,
the recorded EPIC-pn count rate of 
$(9.0\pm2.5) \times 10^{-4}~{\rm cts~s}^{-1}$ corresponds to a
(0.1--10~keV band) luminosity of 
$L_{\rm x}\sim2\times10^{37}~{\rm ergs~s}^{-1}$ for an assumed
power-law spectrum with photon index 2 and a Galactic foreground 
column of $N_{\rm H}=5\times10^{20}~{\rm cm}^{-2}$ \cite{pascual02}.
A similarly low X-ray band luminosity has been inferred for
the Ic SN~1994I (see Sec.~\ref{1994I}).

\newpage
\section{What is so special about X-rays?}
By observing electromagnetic radiation from SNe (and what else is
there in astrophysics, other than neutrinos and gravitational
radiation?), we
can, in principle, obtain information about (i) the nature and mass
of the progenitor, (ii) the possible presence of a companion, (iii)
the history of mass loss and the speed of the progenitor's wind going
back tens of thousands of years, (iv) inhomogeneities in the CSM, (v)
the density gradient of the ejecta as they emerge in the early phase
of the SNe, (vi) the geometry, physical conditions and composition of
the ejected matter, (vii) temperature and density in the region behind
the reverse shock, (viii) speed of the reverse shock as a function of
time, (ix) the geometry, physical conditions and composition of the
CSM, and (x) the speed of the circumstellar shock as a function of
time.

A large part of the story comes from optical observations which
historically have played a key role in our understanding and knowledge
of the various SNe. The classification of the various `types' is
exclusively linked to the optical (see the Chap. by Turatto). A wealth
of information comes directly from the observed IR/optical/UV line
profiles and fluxes (see the Chap. by Branch, Baron \& Jeffery).

The radio emission, which is believed to arise largely from the 
region behind the circumstellar shock, tells its own story (see the
Chap. by Sramek \& Weiler). This story is by and large complementary 
to what we learn from the IR/optical/UV observations. In either
wavelength range, there is not always a unique interpretation that 
explains the complicated spectral behavior. It is therefore perhaps 
not surprising that Filippenko's earlier review \cite{fil97}
is largely phenomenological -- the physics is not always clear, 
though great progress is being made.

The X-ray observations add something special as they preferentially
probe the very hot regions ($T\sim10^{6-10}$~K) and can thus provide 
information elusive in any other wavelength regime.

Very roughly, we can expect three phases of X-ray emission: there
should be a brief burst of high-energy X-rays from Type Ib/c and
II SNe in addition to a black-body continuum of $\sim$0.02 keV as a result 
of the high-temperature flash associated with the break out of the shock
through the stellar surface. In the case of Type Ia, only weak X-ray emission
is expected from the prompt thermal detonation (or deflagration) of the
compact white dwarf. Such brief prompt bursts have not yet been 
observed due to the lack of X-ray all-sky monitors and the response 
time of orbiting X-ray observatories (days).
Weeks or months later, high-energy X-rays may be
detected when the expanding ejecta have become optically thin to
X-rays, and/or when the ejecta plow into CSM supplied by the stellar
wind in previous phases of mass loss of the progenitor. The
circumstellar shock that is formed in this interaction can have a very
high temperature ($T\gtap10^{9}$~K). Radio emission is also believed
to come from the region behind this shock. The reverse shock
is radiative and a dense, cool ($T\ltap10^{4}$~K) shell can form
behind the reverse shock \cite{chevfran94}. If the
density gradient of the outer part of a core collapse supernova is
very large, the density of the reverse shock will be high, and X-rays
from the reverse shock will be heavily absorbed in this dense cool
shell.  When this cool region has expanded sufficiently to become
transparent to X-rays, X-ray emission from the reverse shock will
start to dominate. Radio emission is not expected to come from this
region.  Yet, it is remarkable how well the radio and the X-ray
emission track each other (see Fig.\ \ref{lightcurve}). If there are separate 
shells of CSM as formed in the past (spasmodic bursts of stellar wind from
the progenitor), then the X-ray and radio emission will flare up more
than once.  In the case of SN~1987A the ejecta are just
now beginning to plow into the ring of CSM whose presence was earlier detected
with {\it HST} \cite{burrows95b,burrows00,pod91}.

Immler et al. \cite{immler01} have put a coherent picture together of 
the various phases of X-ray emission in the case of SN~1993J. They were
even able to reconstruct the evolution of the progenitor of SN~1993J,
uncovering $\sim$10$^{4}$ years in wind history, and found
evidence that the progenitor was making a transition from the red 
to the blue supergiant phase during the late stage in the evolution.
This scenario for the evolution of the SN~1993J progenitor revealed 
interesting similarities with that of SN~1987A, whose progenitor 
completely entered the blue supergiant phase after significant mass-transfer 
to a companion (see the Chap. by McCray). The X-ray data of SN~1993J 
cover many years, demonstrating the scientific potential of long-term 
X-ray monitoring of SNe as an important diagnostical tool to probe the 
CSM interaction and the evolution of the progenitor.

In the case of Type IIn SNe (e.g.\ SN~1998S), it may not be easy to
distinguish line emission that comes from the region behind the
reverse shock from line emission that comes from clumpy CSM 
\cite{chugai93}. The clumpy CSM may have been accelerated to $\sim
10{^2}$--$10^{3}~{\rm km~s}^{-1}$, in which case the X-ray lines 
should be Doppler-broadened. 
The \C\ and \X\ observations, however, did not have a high enough
photon statistics to address this question. The \C\ observations of 
SN~1998S, however, did allow for the unprecedented determination of 
the abundances of various elements such as O, Ne, Mg, Al, Si, S, Ar, Ca, 
and Fe \cite{po01a}. Comparing the observed abundance ratios with those 
of the models by Thieleman et al. \cite{thieleman96}, seems to indicate 
that the mass of the progenitor was $\sim18$~M$_{\odot}$. This is the 
first time that a mass determination was made using the X-ray data alone.

\subsection*{Acknowledgments}
We are thankful to Dave Pooley for constructing Fig.\ \ref{lightcurve}
from archival, published and propriety right data of the authors,
and to Roger Chevalier for helpful discussions.

\end{document}